\documentclass[aps, prd, twocolumn,superscriptaddress,preprintnumbers,nofootinbib]{revtex4}

\usepackage{bm,graphicx}
\usepackage{dcolumn}
\usepackage{amssymb}
\usepackage{natbib}

\begin{document}

\newcommand{\fd}{f_{10}}
\newcommand{\As}{A_{\mathrm{s}}}
\newcommand{\At}{A_{\mathrm{t}}}
\newcommand{\ns}{n_{\mathrm{s}}}
\newcommand{\nt}{n_{\mathrm{t}}}
\newcommand{\Obhh}{\Omega_{\mathrm{b}}h^{2}}
\newcommand{\Omhh}{\Omega_{\mathrm{m}}h^{2}}
\newcommand{\Ol}{\Omega_{\Lambda}}
\newcommand{\half}{\frac{1}{2}}
\newcommand{\CMB}{\textsc{CMB }}
\newcommand{\CMBEASY}{\textsc{CMBeasy }}
\newcommand{\TT}{\textsc{tt }}
\newcommand{\TE}{\textsc{te }}
\newcommand{\EE}{\textsc{ee }}
\newcommand{\BB}{\textsc{bb }}
\newcommand{\WMAP}{\textsc{WMAP }}
\newcommand{\UETC}{\textsc{uetc }}
\newcommand{\UETCs}{\textsc{uetc}s}
\newcommand{\ETC}{\textsc{etc }}
\newcommand{\ETCs}{\textsc{etc}s }
\newcommand{\HK}{\textsc{HKP }}
\newcommand{\BBN}{\textsc{BBN }}
\newcommand{\mcmc}{\textsc{MCMC }}
\newcommand{\LL}{\mathcal{L}}
\newcommand{\clover}{\textsc{C}$\ell$\textsc{over}}
\renewcommand{\d}{{\partial}}
\newcommand{\be}{\begin{equation}}
\newcommand{\ee}{\end{equation}}
\newcommand{\bea}{\begin{eqnarray}}
\newcommand{\eea}{\end{eqnarray}}
\newcommand{\mbh}[1]{\textbf{#1}}
\def\lsim{\mathrel{\rlap{\lower4pt\hbox{\hskip1pt$\sim$}}
    \raise1pt\hbox{$<$}}}                
\def\gsim{\mathrel{\rlap{\lower4pt\hbox{\hskip1pt$\sim$}}
    \raise1pt\hbox{$>$}}}                

\preprint{}
\title{Detecting and distinguishing topological defects in future
  data from the CMBPol satellite}   

\newcommand{\addressSussex}{Department of Physics \&
Astronomy, University of Sussex, Brighton, BN1 9QH, United Kingdom}

\author{Pia Mukherjee}
\email{p.mukherjee@sussex.ac.uk}
\affiliation{\addressSussex}
\author{Jon Urrestilla}
\email{jon.urrestilla@ehu.es}
\affiliation{Department of Theoretical Physics, University of the
  Basque Country UPV-EHU, 48040 Bilbao, Spain} 
\affiliation{\addressSussex}
\author{Martin Kunz}
\email{martin.kunz@unige.ch}
\affiliation{D\'epartement de Physique Th\'eorique, Universit\'e de
  Gen\`eve, 1211 Gen\`eve 4, Switzerland} 
\affiliation{\addressSussex}
\author{Andrew R.~Liddle}
\email{a.liddle@sussex.ac.uk}
\affiliation{\addressSussex}

\author{Neil Bevis}
\email{n.bevis@imperial.ac.uk}
\affiliation{Theoretical Physics, Blackett Laboratory, Imperial
  College, London, SW7 2BZ, United Kingdom}

\author{Mark Hindmarsh}
\email{m.b.hindmarsh@sussex.ac.uk}
\affiliation{\addressSussex}

\date{\today}

\begin{abstract}
The proposed CMBPol mission will be able to detect the imprint of
topological defects on the cosmic microwave background (CMB) provided
the contribution is sufficiently strong. We quantify the detection
threshold for cosmic strings and for textures, and analyse the
satellite's ability to distinguish between these different types of
defects. We also assess the level of danger of misidentification of a
defect signature as from the wrong defect type or as an effect of
primordial gravitational waves.  A $0.002$ fractional contribution of
cosmic strings to the CMB temperature spectrum at multipole ten, and similarly a
$0.001$ fractional contribution of textures, can be
detected and correctly identified at the $3\sigma$ level. 
We also confirm that a tensor contribution of $r = 0.0018$ can be 
detected at over $3\sigma$, in agreement with the CMBpol
mission concept study. These results are supported by a model
selection analysis.
\end{abstract}

\maketitle


\section{Introduction}

Cosmological probes are reaching a sensitivity where they are able to
meaningfully constrain models of the early Universe. Data compilations
including Wilkinson Microwave Anisotropy Probe (WMAP) data
\cite{Hinshaw:2008kr,Nolta:2008ih,Dunkley:2008ie,Jarosik:2010iu,Larson:2010gs}
already indicate that the observed inhomogeneities are mostly due to
primordial adiabatic scalar perturbations \cite{Komatsu:2008hk,Komatsu:2010fb}. 
However, there remains room for low-level
contributions from other sources such as cosmic defects
\cite{Wyman:2005tu,Fraisse:2006xc,Battye:2006pk,Bevis:2007gh} and
primordial tensor perturbations, believed to be generated by inflation
alongside the scalars.

These will be detected primarily from the signal they produce in
cosmic microwave background (CMB) polarization, in particular the
B-modes which have yet to be detected and are a target for future
probes.  The possible detection of B-modes produced by primordial
gravitational waves (tensor modes) is often referred to as the
`smoking gun' of inflation.  The amplitude of the primordial
gravitational wave background would provide strong constraints on high-energy 
physics models of inflation, including some appealing models
coming from string theory/brane inflation.  String/M-theory may thus
be constrained by cosmological data: even though there exist models
\cite{McAllister:2008hb} that give rise to a measurable
tensor-to-scalar ratio $r$, a fairly general prediction from string
cosmology seems to be that the level of primordial gravitational
waves, given by $r$, is very low ($r \ll 10^{-3}$, in some cases even
$r\sim 10^{-23}$). As emphasized by Kallosh et
al.~\cite{Kallosh:2007wm} it is hard to obtain an inflationary model
coming from string theory which predicts measurably high primordial
tensor modes. Thus, a future detection of $r$ in the accessible range
$r\gtrsim10^{-2}$--$10^{-3}$ would present important implications for
string cosmology.

Another typical prediction of string cosmology is the production of
cosmic (super)strings
\cite{Majumdar:2002hy,Sarangi:2002yt,Jones:2003da}. Indeed, cosmic
strings are a quite general prediction from high-energy inflationary
models within the Grand Unified Theory (GUT) framework
\cite{Jeannerot:2003qv}.  Strings produced after inflation will also
generate CMB anisotropies
\cite{Vilenkin:1994book,Hindmarsh:1994re,Lyth:1998xn}.  Cosmic strings
are not the only possible cosmic defects in high-energy inflationary
models: global monopoles, semilocal strings and textures are all
examples of cosmic defects that could be created after inflation and
remain consistent with the Universe we observe. Determining 
the nature of cosmic defects would provide invaluable
information on high-energy symmetry breaking.

Defects produce scalar, vector and tensor perturbations.  In contrast
to the standard inflationary model, their vector perturbation modes do
not die out since they are seeded continuously by the defects.  Moreover,
there are no free parameters that quantify the relative amount of
scalar, vector and tensor perturbations independently, only an overall
normalization factor; the relative amount of those perturbations is
fixed for a given model.  As defects produce vector and tensor modes,
they create polarization B-modes directly
\cite{Seljak:2006hi,Pogosian:2007gi,Bevis:2007qz}.  It is interesting
to note that even though cosmic defects can contribute at most a small
fraction of the temperature perturbations, which must be mostly
created by inflationary scalar modes to match the temperature
anisotropy data, they can still be dominant in the B-mode spectra.
Urrestilla et al.~\cite{Urrestilla:2008jv} have shown that Planck
satellite \cite{Planck} data would not suffer from significant
degeneracy between tensors and strings. Thus, if Planck detects
extra ingredients in the B-mode polarization spectra, its accuracy
will be enough to say whether the source of the spectra are primordial
tensor modes or cosmic defects.

CMBPol \cite{Baumann:2008aq} is a proposed space mission that has
higher sensitivity than Planck and is specifically designed to target
the polarization anisotropies. Here we perform an analysis, partly
along similar lines to Ref.~\cite{Urrestilla:2008jv}, to determine
both the detection threshold for different types of signal in CMBpol
data and the ability of the satellite to distinguish between different
defect types as well as primordial tensors. We use both parameter
estimation and Bayesian model selection tools to achieve this. The
detection thresholds we  find improve on those expected from
Planck over an order of magnitude, under realistic
assumptions about foreground residuals and without assuming any level of delensing.

\section{Different cosmic defects.}

High-energy physics models of inflation often give rise to cosmic
defects after inflation ends. The most studied ones are cosmic strings.
These are one-dimensional objects that are extremely long (cosmic
size) and yet microscopic in width, which generate CMB
perturbations. They can arise in field theories (for example, 
they are expected  in SUSY GUT models \cite{Jeannerot:2003qv}) and can also be
present as cosmic superstrings arising  in fundamental string theories
\cite{Majumdar:2002hy,Sarangi:2002yt,Jones:2003da}.

Other kinds of defects are also possible; global defects can be
formed, such as global monopoles or textures
\cite{Pen:1997ae,Durrer:2001cg,Bevis:2004wk,Urrestilla:2007sf}. Global
monopoles do not present a problem in a cosmological setup (contrary
to their local counterparts, the `usual' magnetic
monopoles), because their scaling properties 
are such that their energy density remains a fixed small fraction of the total.
 Textures are also permissable byproducts of
cosmological symmetry breaking processes; indeed textures have been
invoked in the CMB context as a possible explanation of the cold spot 
 \cite{Cruz:2007pe}.

In previous papers by some of us
\cite{Bevis:2004wk,Bevis:2006mj,Bevis:2007qz,Urrestilla:2007sf} we
calculated the CMB power spectra (temperature and polarization
spectra) of cosmic strings, semilocal strings
\cite{Achucarro:1999it,Achucarro:2007sp} and textures from field
theoretical simulations.  There we showed that the spectra from all
these defects are very different from those of primordial inflationary
models (including tensors). We also showed that even though there are
similarities amongst them, there are also differences, with the
semilocal predictions lying somewhere between textures and strings.

\begin{figure}[t]
\hspace*{-0.1cm}
\resizebox{\columnwidth}{!}{\includegraphics{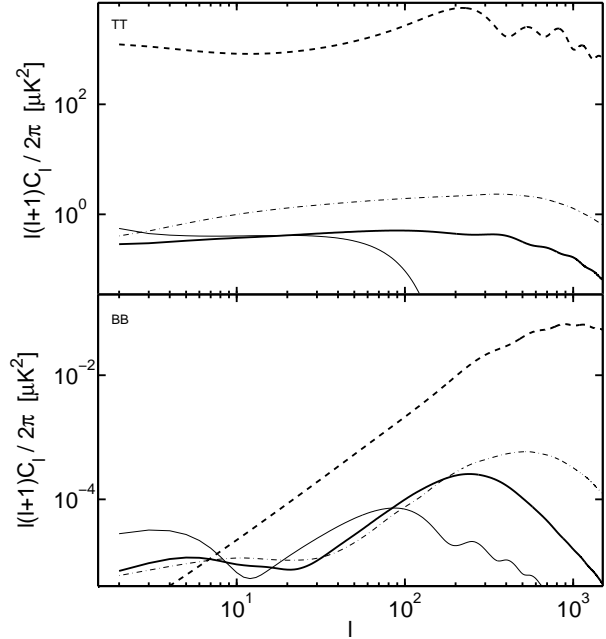}}
\caption{\label{spectra} The CMB temperature and B-mode polarization
  spectra for different components. The figure shows the temperature
  spectra from inflationary scalar modes (black dashed), inflationary
  tensor modes (black thin), cosmic strings (gray dot-dashed), and
  textures (black thick). The normalizations shown are at the
  threshold detection levels identified later in this paper assuming
  the true model is known: for the tensors this is at the $r=0.0018$
  level, for strings $f_{10}^{\rm st} = 0.0012$, and textures $f_{10}^{\rm tex}=
  0.0005$. Clear differences are seen in the shapes of the spectra.}
\end{figure}

The zoo of possible defects is richer than that described here, but
rather than performing an extensive comparison, we choose to focus on
just two of them: cosmic strings and textures. An exhaustive analysis
would not generate further insight at this stage. Besides, the exact
prediction for each kind of defect has its subtleties, and often
different calculational approaches result in slightly differing spectra
\cite{Battye:2010hg}. Our aim is to verify and quantify at what level
the spectra created by two different kinds of defects, such as the
ones shown in Fig.~\ref{spectra}, can be distinguished by CMBpol.

We use the latest, more accurate, spectra derived from field
theoretical simulations of cosmic strings and textures
\cite{Bevis:2010gj}.  Those spectra have been obtained by making the
minimal possible computational changes in order to capture the
differences between those two defect types. The Abelian Higgs model
used to model cosmic strings and the linear $\sigma$-model for the
textures were evolved using the same discretization algorithms, the
same type of initial conditions, and the same procedure to calculate
the power spectra (more details can be found in
Refs.~\cite{Bevis:2006mj,Urrestilla:2007sf}). As in our previous
papers, we quantify the amount of defects by $f_{10}$, which is the
fractional contribution to the total TT power spectrum at
$l=10$. Observational data sets an upper limit on defects of a few
percent: for strings $f_{10}\sim0.1$ \cite{Battye:2006pk,Bevis:2007gh}
and for textures $f_{10}\sim0.16$ \cite{Urrestilla:2007sf}.  In turn,
this parameter $f_{10}$ can be translated into a value of $G\mu$, with
$G$ the gravitational constant, and $\mu$ the string tension. For the Abelian Higgs 
model simulations used in this paper, $f_{10} =0.1$  corresponds to $G\mu \simeq 6\times 10^{-7}$.

For
textures, it is not natural to talk about a ``string'' tension, but we
will use $\mu$ defined as $2\pi \phi_0^2$, where $\phi_0$ is the symmetry-breaking scale, 
to ease comparison (see the appendix of
Ref.~\cite{Urrestilla:2007sf} for more details).

\section{Methods}

We simulate CMBPol data as described in the CMBPol mission concept
study \cite{Baumann:2008aq} in its high-resolution version. The
treatment there follows the approach of Ref.~\cite{Verde:2005ff} in
modelling residuals from foreground subtraction and propagating their
effects into uncertainties in cosmological parameters. 

We consider a flat $\Lambda$CDM model with the same set of fiducial
parameters as used in Ref.~\cite{Baumann:2008aq}: $H_0=72 \, {\rm km \,
  s}^{-1} \, {\rm Mpc}^{-1}$, $\Omega_{\rm b}h^2=0.0227$, $\Omega_{\rm
  c}h^2=0 .1099$, $\tau=0.087$, $A_{\rm s}=2.41\times10^{-9}$, $n_{\rm
  s}=0.963$. In our analysis we vary these parameters in addition to
the tensor-to-scalar ratio $r$ and/or the level of strings and
textures (for which we quote the level of these defects relative to
the total TT power spectrum at multipole $l=10$, which we label as
$f_{10}^{{\rm st},{\rm tex}}$). We assume the inflationary consistency
relation $n_{\rm t}=-r/8$ for the tensor spectral index, and do not
allow running of the scalar spectral index. The inflationary
parameters are specified at a pivot scale of $k _*=0.05 \, {\rm
  Mpc}^{-1}$. We assume that 80\% of the sky can be used for
cosmological analysis.

The effect of lensing in the inflationary spectra is included in the prediction of the signal. We work in the Gaussian limit (ignoring mode correlations due to lensing or defects) where the likelihood takes its usual form \cite{Lewis:2005tp}, and ignore lensing due to defects. The task of detecting defects through B-modes primarily amounts to detecting the excess variance in the $C_l$ from defects against this lensing contribution, which is more or less fixed by the other spectra. Accordingly its recovery is approximately limited by the cosmic variance of the lensing signal, but this is fully modeled through the likelihood. 
We use the pessimistic dust model (third column
of Table 10 of Ref.~\cite{Baumann:2008aq}), and use an intermediate
value for the level of foreground residuals (not 1\% or 10\%, but
5\%).

We simulate instrumental data with the input sources being adiabatic
primordial scalars plus either cosmic strings or textures at different
contribution levels. We then analyze that data in two ways, the first
being a parameter estimation exercise and the second a model
comparison.

We use CosmoMC \cite{Lewis:2002ah} to obtain parameter confidence
contours.  Our fiducial model consists of a flat $\Lambda$CDM model
with the parameters quoted above, and we include some defects (one
case with cosmic strings, the other with textures). Then we try to fit
that simulated data using all the different possibilities that can be
assembled from the different components: a model with strings; a model
with tensors; a model with textures; a model with strings and
textures; with strings and tensors; with textures and tensors; and
with strings, tensors and textures. This exercise allows us to infer
the level of defects needed to clearly distinguish one from the other.

For our model-level analysis, we compute the Bayes factors of the set
of models mentioned above, that is, models with one extra ingredient
(strings, textures or tensors) and models with combinations of two of
those or all three extra ingredients.  In order to obtain the Bayes
factors we use the Savage--Dickey ratio \cite{Verdinelli:1995,Trotta:2005ar},
and we consider two sets of priors for these extra ingredients: flat
linear priors and flat logarithmic priors. The relative Bayes factors
of all these models will pinpoint which of those models is favoured
and at which level.

We have also analysed the data under different assumptions than those
described above. Ignoring foreground uncertainties reduces parameter
error bars by about a factor of two. If we turn off lensing
(i.e. assume that the CMB can be perfectly delensed) then error bars
decrease by a factor of about seven. Thus, an order of magnitude
improvement in $f_{10}$ is still possible over the uncertainties described in the
rest of this paper.

Previous works that considered constraints on strings in CMBPol-like experiments include Refs.~\cite{Seljak:2006hi} 
and  \cite{GarciaBellido:2010if}. However, the assumptions  made there are  different to  those considered here
regarding, for example, polarization sensitivities and lensing residuals.  
In addition, Ref.~\cite{Seljak:2006hi} uses the Unconnected Segment Model \cite{Pogosian:1999np} for string perturbations, 
which has a different $G\mu$ for a given $f_{10}$ unless special parameters are chosen \cite{Battye:2010xz}.

\section{Parameter estimation analysis}

We carry out two kinds of parameter estimation analyses. In the first,
we fit assuming we already know the fiducial model used to generate
the data. This enables us to determine the sensitivity of CMBpol to
the different defect signals, under the best possible circumstances. We
then carry out an analysis in the presence of model uncertainty, to
assess the possible effect of mistaken assumptions.

\subsection{Fitting with the correct model}

If we assume we already know the correct model the analysis is
particularly straightforward.  We first use the fiducial model
described above together with cosmic strings only, with $f_{10}^{{\rm
    st}}=0.0021$ (correspondingly, $G\mu \simeq 9 \times 10^{-8}$).
Fitting for the same parameters as went into the model, strings are
detected at high significance, with $f_{10}^{{\rm st}} = 0.0021\pm
0.0004$. Thus in this best-case scenario, a 0.0012 fractional
contribution from cosmic strings ($G\mu\simeq 7 \times 10^{-8}$)
would qualify for a $3\sigma$ detection, and hence this is the
detection threshold for strings with this CMBpol configuration.

Repeating the analysis with textures, using a fiducial value of
$f_{10}^{{\rm tex}}=0.0007$ ($G\mu \simeq1.2 \times
10^{-7}$), we find that in fitting for textures $f_{10}^{{\rm
    tex}}=0.00070\pm 0.00015$ is obtained. The $3\sigma$ detection
threshold for textures is therefore a 0.0005 fractional contribution
to the CMB TT power spectrum at $l=10$ ($G\mu\simeq 1.0 \times
10^{-7}$).

The power spectra shown in Fig.~\ref{spectra} were normalized to
indicate these detection thresholds. The distinct shapes of the
spectra in both TT and BB are evident, with the defect spectra more
resembling each other than the tensors.

\subsection{Fitting with a range of models}

In reality we do not know {\it a priori} which model is correct, and
indeed our primary interest is likely to be in determining the correct
model. One should be concerned about whether one might be able to draw
conclusions based on the wrong model assumption, e.g.\ in the actual
presence of strings, instead fitting primordial tensors and apparently
detecting $r$ at some significance. We wish also to know whether or
not such data fits are able to draw us towards the correct model
conclusion. The parameter estimation approach of this section is
complemented by the more robust model-level analysis we provide in
Section~\ref{s:model}.

\begin{figure}[t]
\resizebox{0.8\columnwidth}{!}{\includegraphics{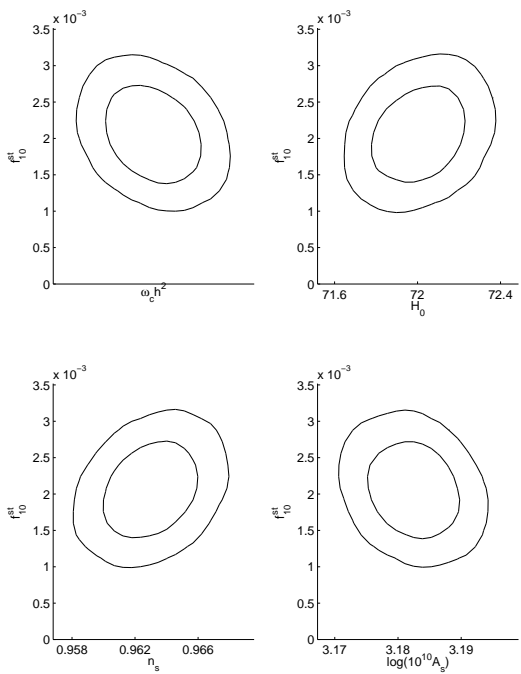}}
\caption{\label{stpt02_s_2D} The correlation between strings and cold
  dark matter density, the hubble parameter, the scalar spectral
  index, and the amplitude of primordial perturbations. These are the
  parameters with which strings are most correlated, at the -37\%,
  34\%, 28\% and -24\% levels respectively.  Textures are less
  correlated with cosmological parameters, the levels being -10\%,
  20\%, 9\%, and -8\% respectively.}
\end{figure}
\subsubsection{True model is strings}

If we fit for tensors instead of strings, we do get a mild detection
of $r$ with $r=0.0012\pm0.0005$, and other parameter recoveries are
biased against their fiducial values; the ones that shift by more than
a sigma are cold dark matter density (which goes up more than
$1\sigma$ to 0.1103), the scalar spectral index (which goes down more
than $1\sigma$ to 0.961) and $H_0$ (goes down more than $1\sigma$ to
71.8).  Parameter correlations are shown in Fig.~\ref{stpt02_s_2D}. If instead we
wrongly fit for textures we get a strong detection of textures with
$f_{10}^{{\rm tex}}= 0.0006\pm0.00015$ and again other parameters get
biased, though less significantly, as expected because textures can
account for part of the string signal (cold dark matter density,
scalar spectral index and $H_0$ shift to 0.1102, 0.962 and 71.8
respectively)  There is therefore a danger of being led astray through
assumption of the incorrect cosmological model. The bias in the values
for other parameters is a potential signal but there may be no
independent means of estimating them to the same accuracy. It is
therefore important to test different model assumptions, and this
motivates attempts to fit multiple components.

\begin{figure}[t]
\resizebox{0.8\columnwidth}{!}{\includegraphics{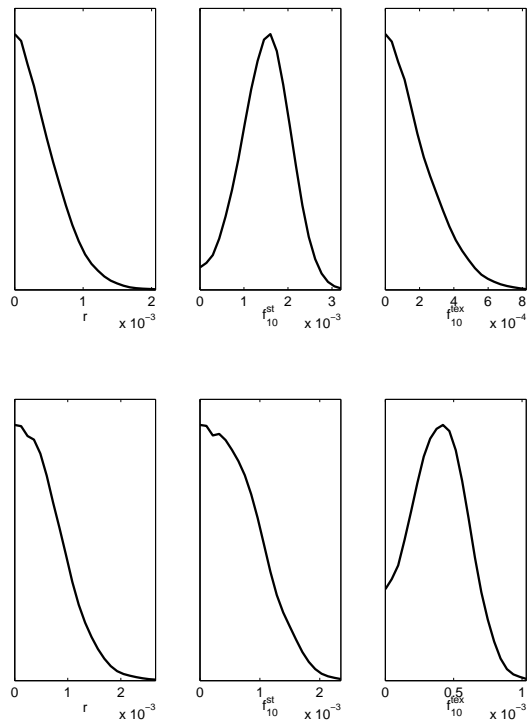}}
\caption{\label{sttex1D} 1D marginalized likelihood for tensors,
  strings, and textures, for the fiducial model with strings
  $f_{10}^{\rm st} = 0.004$ (top row) and for the fiducial model with
  textures $f_{10}^{\rm tex} = 0.0009$ (bottom row). In the texture
  case the peak is not strong enough for even a $2\sigma$ detection.}
\end{figure}

\begin{figure}[t]
\resizebox{0.9\columnwidth}{!}{\includegraphics{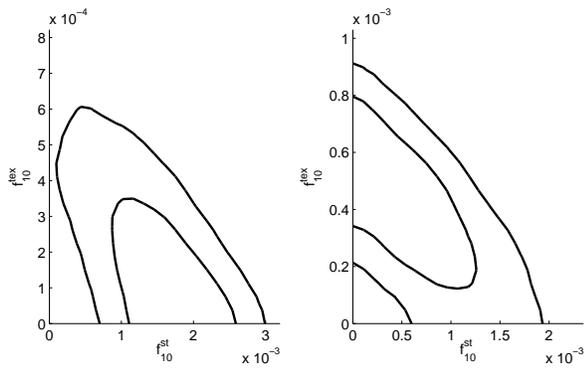}}
\caption{\label{sttex} The correlation between strings and textures in
  simulated CMBPol data with strings (left panel) and textures (right
  panel), showing 68\% and 95\% confidence contours.}
\end{figure}

When fitting for strings and textures together, or for all of strings,
textures and tensors, results are very similar. Figure~\ref{sttex1D}
shows the marginalized likelihoods for each component from a fit where
all parameters are simultaneously varied, the upper panels showing a
fiducial string model.  From these likelihoods we get $f_{10}^{{\rm
    st}}=0.0015\pm 0.0005$, while $f_{10}^{{\rm tex}} < 0.0002 \mbox{
  (68\% c.l.)}, 0.0005\mbox{ (95\% c.l.)}$ and $r<0.0005\mbox{ (68\%
  c.l.)}, 0.0011\mbox{ (95\% c.l.)}$ receive only upper limits.  The
level of correlation between strings and textures is found to be
$\approx 60\%$, shown in Fig.~\ref{sttex}. Tensors are not much
correlated with strings (10\%) or textures (15\%) (Fig.~\ref{spectra}
shows that strings peak at roughly twice as high an $l$ as textures,
and at low $l$ strings have a little peak where the tensors dip,
making strings a little more dissimilar to tensors than
textures). Table~\ref{threecomp} summarizes the uncertainties found
under various assumptions.

The correct component can also be sought based on the quality of the
fits, i.e.\ by looking at the best-fit and mean likelihoods achieved
in the MCMC analysis. Strings+textures+tensors and strings alone lead
to similar best-fit or mean log-likelihood, in each case being at
least 3 better than with textures alone and 10 better than tensors
alone. We conclude that at this fiducial string contribution, the
model of tensors alone could be discounted, and strings favoured over
textures, though not convincingly.

We find that a slightly higher fiducial value of $f_{10}^{{\rm
    st}}=0.003$ gives a $4\sigma$ detection ($f_{10}^{{\rm
    st}}=0.0024\pm0.0006$), in a joint fit, while the other components
continue to receive upper limits. The recovered string fraction is
underestimated because part of the string signal gets ascribed to
textures in the fits. Such results would be a strong indication that
strings were the right model, and a subsequent refit varying the
string amplitude alone would remove this recovery bias. The $3\sigma$
threshold for identifying strings correctly in favour of these
alternatives is therefore $f_{10}^{\rm st} \simeq 0.002$ ($G\mu
\simeq9 \times 10^{-8}$).

\begin{table}[t]
\begin{ruledtabular}
\begin{tabular}{c|ccc}
Model has& $\delta f_{10}^{{\rm st}} $ & $\delta f_{10}^{{\rm
        tex}}$ & $\delta r$ \\ 
\hline \hline
String & $0.00041$ & $-$ & $-$ \\ 
\hline
String & $-$ &$0.00015$ & $-$ \\ 
String & $-$ &$-$ & $0.00052$\\ 
\hline
String & $0.00056$ & $0.00026^*$ & $-$ \\ 
\hline
String & $0.00055$ & $0.00025^*$ & $0.00055^*$ \\ 
\hline \hline
Texture & $-$ & $0.00015$ & $-$ \\ 
\hline 
Texture& $0.00041$ & $-$ & $-$ \\ 
Texture& $-$ & $-$ & $0.00054$\\ 
\hline 
Texture & $0.00071^*$ & $0.00019$ &$-$\\ 
\hline
Texture & $0.00067^*$ & $0.00023^*$ & $0.00070^*$ \\
\end{tabular}
\end{ruledtabular}
\caption{\label{threecomp} Standard deviation achieved when trying to
  fit the data with a model with one, two or three extra
  components. In the string case the fiducial value is $f_{10}^{\rm
    st} = 0.0021$, and for textures $f_{10}^{\rm tex} = 0.0007$. In
  each block of five, the first section corresponds to fitting with
  the correct component, the second and third the wrong component, and
  the fourth and fifth fitting multiple components including the
  correct one. The stars (*) denote the cases when only upper limits are 
placed and the numbers
  quoted are the difference between the 68\% and 95\% upper limits. (Just 
in this table we quote accuracies to an additional significant figure as 
compared to the rest of the text.)}
\end{table}

\subsubsection{True model is textures}

If we fit for tensors instead of textures, we get a false detection
$r=0.0016\pm 0.0005$, while other parameters do not get much biased
away from their input values. Biases are smaller because textures are
less correlated with cosmological parameters (see caption of 
Fig.~\ref{stpt02_s_2D}). If we perform the fit for strings instead of textures, strings
receive a false detection with $f_{10}^{{\rm st}}=0.0017\pm 0.0004$,
with very insignificant parameter shifts this time in the opposite
direction.  Thus in both cases a false detection of the wrong
component occurs for the level of textures assumed in the fiducial
model.

When all of textures, strings and tensors are fitted for, then the
input level of textures proves to be too low for a clear
detection. Instead all components receive upper limits: $f_{10}^{{\rm
    tex}} <0.0005\mbox{ (68\% c.l.)}, 0.0007 \mbox{ (95\% c.l.)}$,
$f_{10}^{{\rm st}} < 0.0008\mbox{ (68\% c.l.)}, 0.0015\mbox{ (95\%
  c.l.)}$ and $r < 0.0008\mbox{ (68\% c.l.)}, 0.0015\mbox{ (95\%
  c.l.)}$; see the lower panels of Fig.~\ref{sttex1D}. Thus a 0.0007
contribution of textures to the CMB TT power spectrum at $l=10$ is not
clearly detectable when fitting for these two additional parameters.

Study of best-fit and mean likelihoods will again enable a ranking of
models considered, with models involving textures preferred by about 3
(6) as compared to models with just strings (tensors).

Further simulations show that $f_{10}^{{\rm tex}}=0.0010$ in the data
gives a $3\sigma$ detection of textures ($f_{10}^{{\rm
    tex}}=0.0006\pm0.0002$) and $f_{10}^{{\rm tex}}=0.0012$ gives a
$4\sigma$ detection of textures ($f_{10}^{{\rm
    tex}}=0.0009\pm0.0002$), in the case when all components are
fitted. As above, a part of the texture signal gets ascribed to
strings in the fits, reducing the recovered texture signal when
strings are also allowed. The threshold for identifying textures
correctly in favour of these alternatives thus is $f_{10}^{\rm tex}
\simeq 0.001$ ($G\mu \simeq 1.4 \times 10^{-7}$).

\section{Model selection analysis}
\label{s:model}

Until we have uncovered the presence of defects, we are less
interested in constraints on the defect parameters and more in the
fundamental question whether there are any defects in the Universe,
and if yes, which kind. This is a question of model selection rather
than parameter estimation, and should be dealt with by computing Bayes
factors between the different models, including a model with no extra
ingredients.\footnote{In this section we will refer to the
  model with no extra ingredients as the `no defect' model, meaning a
  model without strings or textures, but also without tensors.} In
this section we will only consider the fiducial string model.

Before embarking on a model selection analysis, we need to consider
the priors that we want to place on the parameters. Here we will look
at two different priors. In the first one, we assume that the prior is
uniform in $f_{10}$ in the interval $[0,1]$ for all extra
contributions. This interval is much wider than the precision of
CMBPol, which will lead to a significant ``Occam's razor'' factor that
the defect models need to overcome in order to be favoured against the
no-defect baseline model. While this prior makes sense when looking
for a signal that may be present at some level in the $C_\ell$, it
appears at least as natural to impose a prior which assumes that the
phase transition in which the defects were generated happens with
equal probability at an arbitrary energy scales with some
cut-off. This leads to a prior that is uniform in $\log_{10} G \mu$ for
cosmic strings, or more generally uniform in the logarithm of the
amplitude. As limits we choose that the energy scales
($\sim\sqrt{\mu}$) would range from the GUT scale ($\sim10^{16}$ GeV)
to the SUSY breaking scale ($\sim10^3$ GeV). This in turns translates
into values of $\log_{10} f_{10}$ ranging from $-52$ (SUSY scale) to 0 (GUT
scale). Actually, $\log_{10} f_{10}=0$ corresponds to a situation where all
the CMB signal is coming directly from strings, and it is the absolute
maximum number possible according to the definition of $f_{10}$. As we
show in the Appendix, our Bayes factors are only slightly changed when
choosing other physical scales for our lower cut-off for the prior.
For $r$ there is not quite an equivalent to the symmetry breaking
scales of the universe since it is a ratio of tensors to scalars
rather than on an absolute scale, so we just used the same range as
for $f_{10}$, i.e., $r$ ranges from 1 to $10^{-52}$ in our logarithmic
prior analysis.

Having specified the priors, we are left with the technical question
of how to compute the Bayes factors. One possibility is to compute the
model probabilities directly using for example nested sampling
\cite{skilling:395,Mukherjee:2005wg,Bassett:2004wz}. Here we instead
employ the Savage--Dickey (SD) density ratio
\cite{Verdinelli:1995,Trotta:2005ar}, since a model with a given kind of
defects is nested within the simpler model without defects at the
point $f_{10}=0$. The Bayes factor in favour of the simpler model is
then just the value of the (marginalized and normalized) posterior at
$f_{10}=0$ divided by the prior at the same point. For our linear
prior in the defect amplitude, the prior is always just equal to
$1$. We include an Appendix in which we describe the techniques
employed in order to accurately obtain the normalized posterior values
and subsequent Bayes factors.

For our analysis, we ran chains with a fiducial cosmic string
fractional amplitude of $f_{10}^{{\rm st}}=0.0021$ and for models that
included only one kind of extra contribution (strings `s', texture `t'
and tensor modes `r'), two kinds (`st', `sr' and `tr') and all three
(`str'). We then used the SD ratio to derive the Bayes factors with
respect to all nested models, e.g. `st' to `s' and `t'. In this way we
built a partially redundant tree of model probabilities, starting with
the basic model of `no defects' which we used as the reference for
defining relative probabilities. The partial redundancy allowed us to check
whether the results from different paths through the model space are
consistent: it is for example possible to reach the `st' model through
the sequence `no defects' $\rightarrow$ `s' $\rightarrow$ `st' as well
as through `no defects' $\rightarrow$ `t' $\rightarrow$ `st'. The
Bayes factor from both sequences must agree within the error
bars. Indeed this is the case for all results quoted in the paper.

\subsection{Analysis for linear prior}

\begin{figure*}[t]
\resizebox{\columnwidth}{!}{\includegraphics{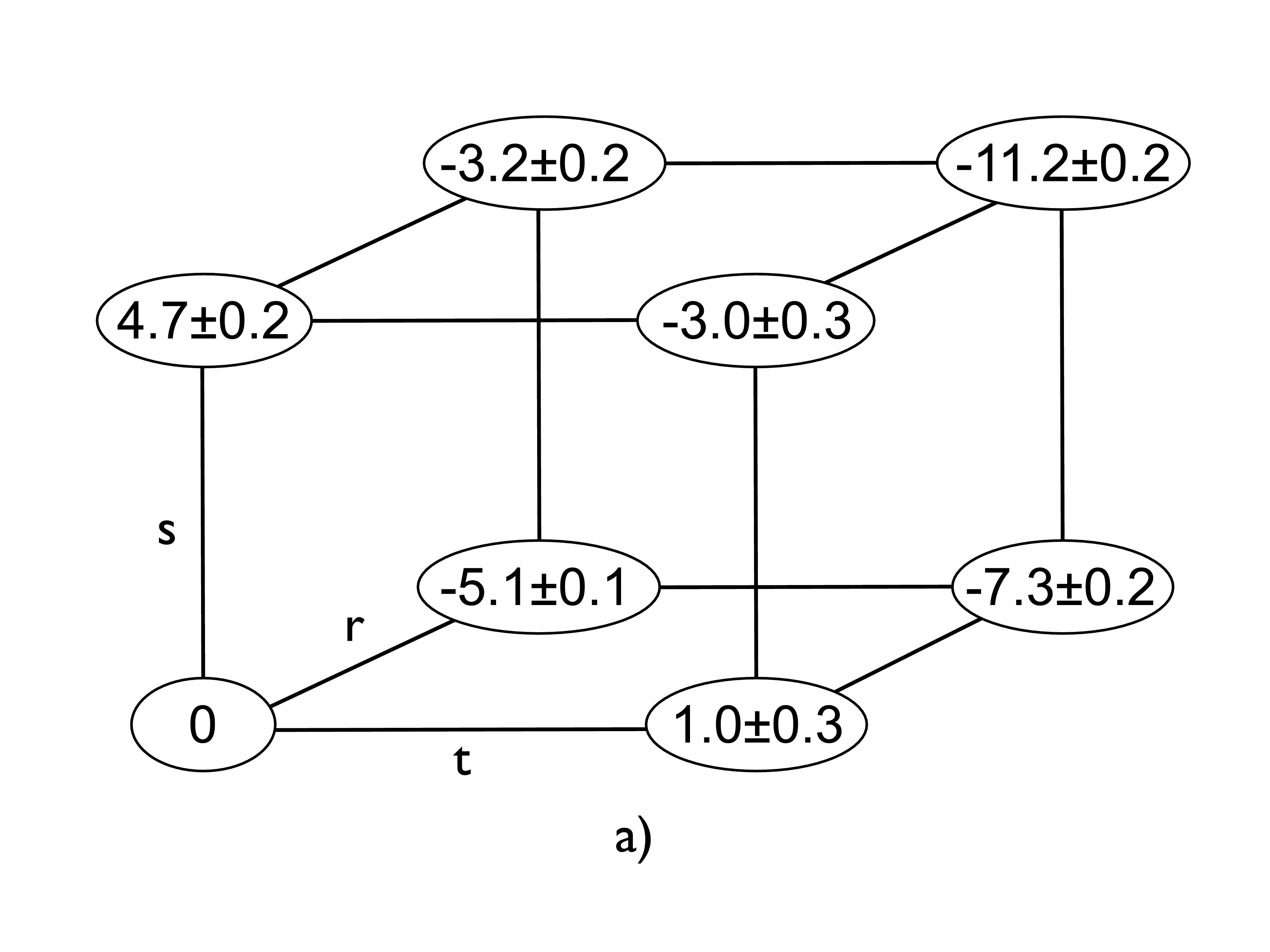}}
\resizebox{\columnwidth}{!}{\includegraphics{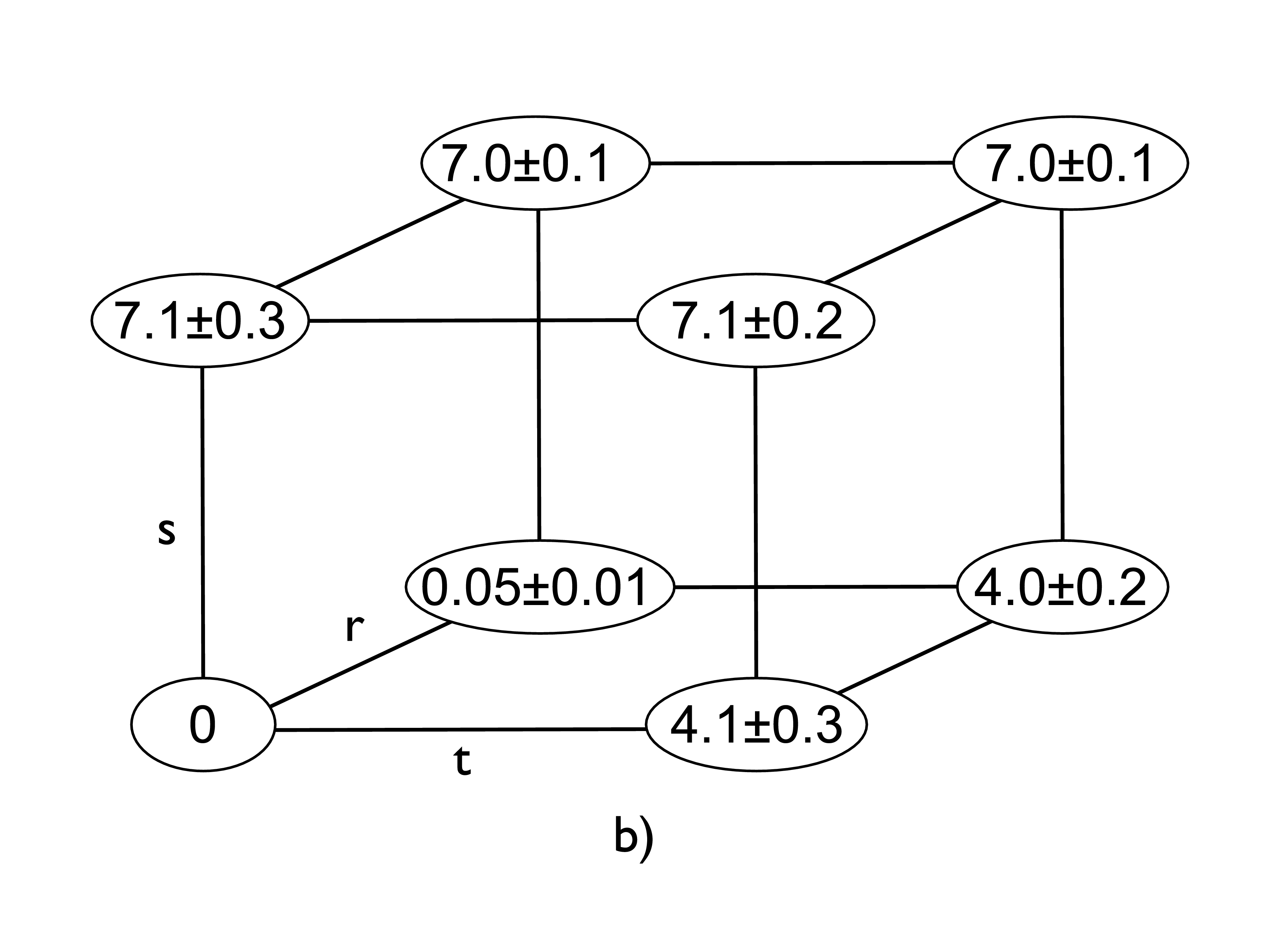}}
\vspace{-0.8cm} 
\caption{\label{cube} Pictorial representation of the logarithm of the
  Bayes factors for different models, relative to a model with `no
  defect'. The lower left corner of the cube corresponds to the `no
  defect' model, and the axes of the cube correspond to adding strings
  (s), textures (t) or tensors (r). Thus, the diagonally opposite
  corner correspons to a model with strings+textures+tensors. Figure a) depicts  a model
  comparison with linear priors in $f_{10}$, whereas in figure b) we 
have used a logarithmic prior instead.}
\end{figure*}

We first performed the analysis described above using linear priors
for all the three extra components.  Figure~\ref{cube}a shows the
results from this exercise, in the shape of a cube with each axis
denoting the presence of strings, textures and tensors
respectively. The model in the lower left corner of the cube is the
`no defect' model while the diagonally opposite corner corresponds to
`str'. The numbers given denote $\ln B$, with positive values for
models that are favoured over the `no defect' case.  The fiducial
string amplitude was chosen so that there is strong evidence for the
presence of strings, which means that it is difficult to evaluate the
SD ratio far out in the tail of the distribution in the `s' case.  The
Appendix discusses how we evaluated SD ratios for these cases. We
notice that the only other model with a positive evidence is `t',
which is due to the partial degeneracy of the strings and textures. The `r', `rt' and `rts' models are significantly disfavoured.

Given these results we would conclude that there is strong evidence in
favour of defects -- indeed, this is about the minimal string
contribution for which CMBPol would be able to make such a
statement. We notice that in a parameter estimation context, the
significance is $5\sigma$ if we only fit for the {\it correct}
component (supposing we know which that component is), and it is a
borderline detection when fitting for all three components (of order
$3\sigma$).

As in the parameter estimation case, we would still not be able to
distinguish decisively between strings and texture, although strings
are favoured by a factor of roughly 40 ($\Delta\ln B = 3.7$). To be
able to pinpoint strings as the origin of the observed signal would
require either a larger defect amplitude or else a more sensitive
probe.

\subsection{Analysis for logarithmic prior}

We also performed a model selection analysis using logarithmic priors
for the extra parameters. Obtaining SD ratios for this case is even
harder than for the linear case, since it presents the additional
difficulty that the models are not actually nested, since $f_{10}=0$
is not attainable (recall that the range for our priors is $\log_{10}
f_{10}\in[-52,0]$). The Appendix deals with the techniques used to
obtain the SD ratios, and in this section we just present and discuss
the results.

The outcome of this exercise should be very different to the one
obtained with linear priors: for any prior range, the vast majority of
the priors are always equivalent to the no-defect model. Therefore it
is much harder to rule out defects if they are wrong as they will make
the same predictions as no-defect in too much of their prior space. We
can only get a signal if in some part of those prior spaces we can get
a much better fit, which will enable us to rule out the
no-defect case if there are defects.

We thus expect models without strings to all  be indistinguishable
from each other, and the models with strings to all be
indistinguishable from each other. In fact, this is roughly what we
see in Fig.~\ref{cube}b, where the results for the logarithmic prior
case are shown. As it can be seen, all models containing strings are
now strongly favoured ($\ln B \approx 7$). Tensors alone have a Bayes
factor similar to the no-defect model. However, models with texture
but no strings also get support ($\ln B \approx 4$), due to the
degeneracy between strings and textures. This  allows us to
conclude that models with defects (strings or textures) are actually
favored with respect to the no-defect model. However, as in the linear
case, it is not conclusive in distinguishing between strings and
textures, even though models with strings have the highest Bayes
factors.

\section{Conclusions}

Simulating data as per the CMBPol mission concept study, propagating
uncertainties due to foreground residuals, we find that the level of
cosmic strings and textures that can be detected and correctly
identified (at $3\sigma$) by CMBPol is 0.002 and 0.001 of the total TT
power spectrum at multipole 10 respectively (correspondingly, $G\mu
\simeq 9 \times 10^{-8}$ for strings and $G\mu \simeq 1.4 \times
10^{-7}$ for textures). Similarly a tensor fraction of 0.0018 should
be discernible. Contributions from strings and textures are highly
correlated with each other, so at lower levels the signal would be
harder to attribute to one or the other conclusively. Tensors are not
much correlated with strings but are somewhat correlated with
textures.

We also performed a model selection analysis for a fiducial model
containing cosmic strings. Using a flat prior on $f_{10}$, we found
that a model with only strings is favored over all models but the
texture-only model is also better than a model without any
defects. Models with several types of defects are all strongly
disfavored because of a large ``Occam's razor'' factor. This changes
when taking a prior that is flat in $\log_{10}(G\mu)$.  In this case, it is
not possible to rule out the presence of defects, and all models
containing strings are strongly favoured (with models containing
textures but no strings having an intermediate probability).

A CMBPol-like experiment, as has been proposed both in the US and Europe,
 has the ability to illuminate us on important
issues regarding high-energy physics in the early universe, that we
can only speculate about at this time. It is roughly two orders of
magnitude better (in $f_{10}$) than what we can achieve at this point with WMAP, and over 
an order of magnitude
better than what the
Planck mission will achieve.

\appendix
\section{Posterior calculation for the Savage--Dickey density ratio}

In this appendix we summarise the different techniques used in this
work to perform the model selection analysis. In both prior choices
(linear and logarithmic) for the defect contribution, we encounter
challenges in obtaining an accurate normalized posterior. We will explain
those challenges and describe how we overcame them.

For both prior choices, the interval for the parameter characterizing
the string contribution is much wider than the precision of CMBPol,
which leads to a significant ``Occam's razor'' factor. For example, if
the posterior was Gaussian with a variance of $\sigma^2$ and the prior
flat with a width of $1$, then its normalization alone contributes a
factor
\be
\frac{1}{\sqrt{2 \pi \sigma^2}} \gsim 300
\ee
in favour of the simpler no-strings model, where we used the variance
of the string contribution discussed earlier.  In order to strongly
support the presence of strings, we need to overcome this factor as
well as reach down to at least $\exp(-5)\approx 1/150$ with the
posterior. We therefore need to have an accurate estimate of the
posterior over four to five orders of magnitude. A normal MCMC chain
would need to be exceedingly long to reach that far out; we would
effectively be counting only every 50,000-th sample! This problem can
be alleviated by running MC chains at higher temperatures (we used
$T=2$, and $T=4$ where necessary) that probe the tails much
better. This increases the computational cost of using the SD ratio,
but not prohibitively, especially since we found that scaled versions
of the $T=1$ covariance matrix were sufficient to use for the proposal
densities of the higher-temperature chains. For $T\neq 1$ we are not
sampling from the desired probability density but from $\exp(-
\lambda/T)$ where $\lambda = -\ln \LL$ with $\LL$ is the
likelihood. If we scaled the Gaussian proposal distribution the same
way, then the covariance matrix $C_1$ at temperature $T_1$ should be
changed to $C_2 = C_1 (T_2/T_1)$. What we did in practice was to
increase the proposal scale in CosmoMC from $2.4$ to $3.6$ whenever we
doubled the temperature, i.e. an increase by a factor of $1.5$ (close
to the theoretical value of $\sqrt{2}$), which worked very well.

\begin{figure}[t]
\resizebox{\columnwidth}{!}{\includegraphics{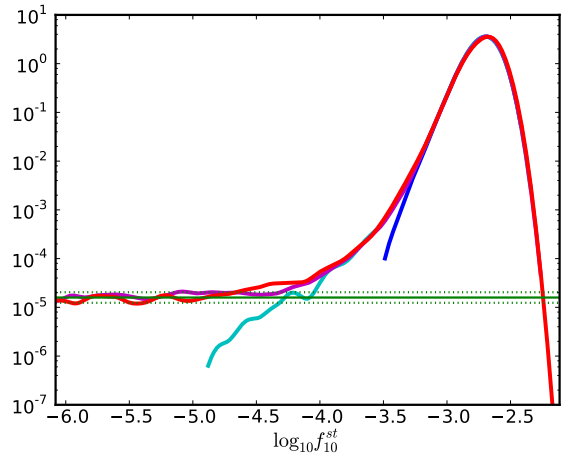}}
\caption{\label{fig:sd_temp} The marginalized probability distribution
  function for $f_{10}^{\rm st}$ in the logarithmic prior case for four
  different temperatures: $T=1$ (blue), $T=2$ (cyan), $T=4$ (magenta)
  and $T=8$ (red) [the curves with lower temperatures end at higher $f_{10}^{\rm st}$]. 
  All chains agree in the high-probability region
  near $\log_{10}f_{10}^{\rm st} = -2.7$, but only the two highest
  temperature chains can probe the low-probability tail which is
  reached for $f_{10}^{\rm st} \rightarrow 0$. }
\end{figure}

In order to use a chain with $T\neq 1$ for model selection and
parameter estimation, we need to correct for the temperature.  This
can be done through importance sampling, by adjusting the sample
weights $w_i$,
\be
w_i(T=1) = w_i(T) \frac{\LL(T=1)}{\LL(T)} = w_i(T)
\frac{e^{-\lambda}}{e^{-\lambda/T}} . 
\ee
Figure~\ref{fig:sd_temp} shows four chains for different temperatures
that use the correction given above.  The resulting probability
densities agree well in the high-probability peak, but the high-$T$
chains probe the low-probability regions much better.

The logarithmic prior presents the additional difficulty that the
models are not actually nested since $f_{10}=0$ is not
attainable. However, we know from the previous discussion about
parameter constraints that, e.g., a defect fraction of $f_{10}=10^{-6}$
is completely undetectable by CMBPol and corresponds for all practical
(although maybe not for philosophical) purposes to a model without
defects. For this reason, the Bayes factor of the model with
$f_{10}=10^{-6}$ relative to the more general model with arbitrary
defect contribution is the same as the one of a model with
$f_{10}=0$. But the former model {\em is} nested in the general model
and allows us to compute the Bayes factor with the help of the SD ratio.

Additionally we do not want to sample all the way down to very big
negative values of $\log_{10} f_{10}$ in the chains (in our case
$\log_{10} f_{10}=-52$), since we already know that the posterior will
be flat once we are below the detection threshold for the defects (see
Fig.~\ref{fig:sd_temp}).  For this reason we impose a cut-off for the
chains at a not so tiny value of $f_{10}$ (in the example described in
the main text, we used $f_{10}=10^{-6}$).  This cut-off is arbitrary
as long as it is in the asymptotic region where the posterior has
become flat since defects are no longer detected. In addition, it is
better to choose it slightly lower than the value at which we evaluate
the SD ratio in order to avoid edge effects.

We proceed as follows: first we evaluate the posterior at the chosen
point (in our example, $\log_{10} f_{10} = -6$). Note that the value
obtained is normalized to the width of the prior actually used in the
calculation (in $\log_{10}$ units, the width is $\epsilon = 6$). Since
we are in the region where the posterior is already flat, we add a
stretch of width $\Delta$ (in our example $\Delta=46$ to reach
$\log_{10} f_{10}= -52$), but we have to be careful with
normalization, since the posterior has now to be normalized taking
into account this new stretch.  Let us denote by $q$ the value
measured at the cut-off point inside the asymptotic region
(e.g.\ $q=1.6\times 10^{-5}$ in Fig.~\ref{fig:sd_temp}). Then, the
normalized posterior actually has a value of $p=q/(1+q \Delta)$. The
prior is flat over the whole range so that its value is
$1/(\Delta+\epsilon)$ and hence the Bayes factor is
\be
B = \frac{1+q \Delta}{q(\Delta+\epsilon)} .
\label{b}
\ee

It is worth checking how much the Bayes factor changes for different
values $\Delta$. Recall that $\Delta$ basically gives us the lower
energy scale taken into account in the prior for
$\log_{10}f_{10}$. For example, $\Delta=46$ corresponds to SUSY
breaking scale, and $\Delta=50$ would correspond to the electroweak
scale.  It is easy to verify that $\ln B(\Delta=46)\sim 7.09 $ and
$\ln B(\Delta=50)\sim7.01 $, so a wide range of lower cutoffs would
give virtually the same results.

Unfortunately there is one additional hurdle that needs to be
overcome. In the model analysis we have performed, we often need to
fit for several kinds of components: strings (s), textures (t) and
tensors (r); and combinations of them: `st', `sr', `tr' and `str'
models. In order to compute SD ratios for models with more than one
component, we marginalize over all but one component.  However, this
marginalization should be done not over the limited range of $f$ that
we actually sample from, but over the full range (taking account the
stretch $\Delta$ in all extra components).

We illustrate how this was performed in the concrete example `st'
$\rightarrow$ `s', where the simulated range of the parameters is
smaller than the full range both in $f_{10}^{\rm st}$ and in $f_{10}^{\rm tex}$.
 First we get the weight of the full simulated chain $W_{\rm
  st}^{\rm sim}$, and the (normalized) pdf $p_{\rm st}^{\rm sim}$ of
the `st' chain by marginalizing over all parameters except $f_{10}^{\rm tex}$.
 That would be enough if our model was nicely nested and we
did not have to add the stretch $\Delta$.  To account for the
stretch, we consider the interval of unit width $\log_{10} f_{10}^{\rm st}
\in [-6,-5]$ of the `st' chain. We calculate both its weight $W_{\rm
  st}^{(1)}$ and its normalized pdf $p_{\rm st}^{(1)}$, once again by
marginalizing over all parameters except $f_{10}^{\rm tex}$.  Note that
$p_{\rm st}^{(1)}$ should be the same as the pdf obtained from the `t'
chains, since $p_{\rm st}^{(1)}$ is calculated virtually in the region
with only textures (no strings). We verified that this was the case.

We now have all the ingredients that are necessary to marginalize over
the full range: Let $W_{\rm st}^\Delta= \Delta\times W_{\rm st}^{(1)}
$ be the estimated weight over all the $\Delta$ stretch. Then the
normalized marginalized pdf for the full range is given by
\be
p_{\rm st}^{\rm full}= \frac{p_{\rm st}^{(1)} W_{\rm st}^\Delta +
  p_{\rm st}^{\rm sim} W_{\rm st}^{\rm sim}}{ 
W_{\rm st}^\Delta + W_{\rm st}^{\rm sim}}
\ee
We still have to account for the $\Delta$ stretch in $f_{10}^{\rm tex}$,
but we are just in the case tackled earlier in this section, and one
just needs to apply Eq.~(\ref{b}) to get the Bayes factor.\\

\begin{acknowledgments}

We are grateful for useful discussions with Jos\'e Juan Blanco-Pillado, Julian
Borrill, Anthony Challinor, Marina Cortes, Stacey-Jo Dias, Juan Garc\'\i a-Bellido and Antony Lewis.  We
acknowledge support from the Science and Technology Facilities Council
[grant number ST/F002858/1] and the Swiss NSF (M.K.). J.U. and
M.K. thank the Benasque Centre for Science, where part of this work
was done, for hospitality. This work was partially supported by the
Basque Government (IT-559-10), the Spanish Ministry FPA2009-10612, and the
the Spanish Consolider-Ingenio 2010 Programme CPAN (CSD2007-00042)
(J.U.). The analysis was performed on the Archimedes cluster of the
University of Sussex and the Andromeda cluster of the University of
Geneva.

\end{acknowledgments}

\bibliography{references}

\end{document}